\documentclass[epsfig,12pt]{article}
\usepackage{epsfig}
\usepackage{graphicx}
\usepackage{array}
\usepackage{bm}
\usepackage{cite}
\usepackage{geometry}
\usepackage{latexsym}
 \usepackage{amsmath, amssymb, amscd, xypic, graphicx}

\newcommand{\beq}{\begin{equation}}   
\newcommand{\eeq}{\end{equation}}
\newcommand{\be}{\begin{equation}}   
\newcommand{\ee}{\end{equation}}
\newcommand{\beqn}{\begin{eqnarray}}   
\newcommand{\eeqn}{\end{eqnarray}}

\newcommand*\xbar[1]{%
 \kern0.5ex%
  \hbox{%
   \kern0.2ex%
      \vbox{%
      \hrule height 0.5pt % The actual bar
      \kern0.5ex%         % Distance between bar and symbol
      \hbox{%
        \kern-0.1em%      % Shortening on the left side
        \ensuremath{#1}%
        \kern-0.1em%      % Shortening on the right side
      }%
    }%
  }%
}

\newcommand{\gsim}{\lower.7ex\hbox{$
\;\stackrel{\textstyle>}{\sim}\;$}}
\newcommand{\lsim}{\lower.7ex\hbox{$
\;\stackrel{\textstyle<}{\sim}\;$}}
\input{my_opt.sty}

\begin{document}

\title{
\begin{flushright}
{\small {IFUP-TH-2017,  FTPI-MINN-17/23, UMN-TH-3707-17 }} 
\end{flushright}
\vspace{5mm}
{  \Large \bf  
Patterns of Symmetry Breaking in Chiral ``QCD"}
 }
\vspace{6mm}
\author{  Stefano Bolognesi$^{(1,2)}$, 
 Kenichi Konishi$^{(1,2)}$, Mikhail Shifman$^{(3,4)}$    \\[13pt]
{\em \footnotesize
$^{(1)}$Department of Physics ``E. Fermi", University of Pisa}\\[-5pt]
{\em \footnotesize
Largo Pontecorvo, 3, Ed. C, 56127 Pisa, Italy}\\[2pt]
{\em \footnotesize
$^{(2)}$INFN, Sezione di Pisa,    
Largo Pontecorvo, 3, Ed. C, 56127 Pisa, Italy}\\[2pt]
{\em \footnotesize  $^{(3)}$  William I. Fine Theoretical Physics Institute,
University of Minnesota,}\\[-5pt]
{\em \footnotesize
Minneapolis, MN 55455, USA}\\[2pt]
{\em \footnotesize  $^{(4)}$ Kavli Institute for Theoretical Physics, UC Santa Barbara, CA 93106, USA
}
\\[3pt]
{ \footnotesize  stefano.bolognesi@unipi.it, \ \  kenichi.konishi@unipi.it,  \ \  shifman@umn.edu}  
}

\date{February, 2018}

\maketitle

\vspace{-5mm}

\begin{center}
{\large\bf Abstract}
\end{center}

 We consider $SU(N)$ Yang-Mills theory with massless chiral fermions in a complex representation of the gauge group. 
 The main emphasis is on the so-called hybrid $\psi\chi\eta$ model.
 The possible patterns of realization of the continuous chiral flavor symmetry are discussed. We argue that the chiral symmetry
is broken in conjunction with a dynamical Higgsing of the gauge group (complete or partial) by bi-fermion condensates. As a result a color-flavor locked symmetry is preserved. The 't Hooft anomaly matching proceeds via saturation of triangles by massless composite fermions or,  in a mixed mode, i.e. 
also by the ``weakly'' coupled fermions associated with dynamical Abelianization, 
supplemented by a number of Nambu-Goldstone mesons. 
Gauge-singlet condensates are of the multifermion type and, though it cannot be excluded, the chiral symmetry realization 
via such gauge invariant condensates is more  contrived (requires a number of four-fermion condensates simultaneously and, even so, problems remain),
and less plausible. We conclude that in the model at hand, chiral flavor symmetry implies dynamical Higgsing by bifermion condensates.

\vspace{2cm}

\newpage

\section{Introduction and  discussion}

The pattern of the chiral symmetry breaking ($\chi$SB) in QCD with $N_f$ massless Dirac flavors is well-known.
$\chi$SB occurs at strong coupling. The main tool for its analysis is matching the 't Hooft triangles \cite{th}, combined with the large-$N$ 
limit and the fact that the singlet axial current is internally anomalous. Applying these tools one concludes
that the global symmetry $SU_f(N)\times SU_f(N)\times U(1)$ is broken down to the vector subgroup $SU_f(N)\times U(1)$
with $N_f^2-1$ Nambu-Goldstone  (NG)  particles which saturate 't Hooft's anomaly matching conditions.

Some models with the chiral fermions were explored in the past  \cite{Raby,Shrock,Poppitz}. 
We will focus on a $G_c=SU(N)$ gauge theory with Weyl fermions $\Psi$  in a complex representation of $G_c$, and without scalars. 
The main emphasis is put on the so-called hybrid $\psi\chi\eta$ model suggested 2012 in \cite{AS} which was argued to be planar equivalent to super-Yang-Mills. 
The pattern of $\chi$SB in this model has not been studied previously as well as its interrelation with the planar equivalence.
We will also comment on the $\psi\eta$ and $\chi\eta$ models discussed previously \cite{Shrock,Poppitz} in the context of the Fradkin-Shenker continuity
\cite{Fradkin},  for a recent perspective see \cite{shify}. This continuity states that  if the order parameter is the vacuum expectation value (VEV) of an operator in the fundamental
representation of the gauge group, then there are no separate Higgs or confinement phases. A single Higgs/confinement phase  exists in this case,
with a crossover from a Higgs picture to the confinement picture, with no phase transition.
In fact, the Fradkin-Shenker concept requires a qualification which is much less known than the concept itself. This is the reason why the $\chi\eta$ model
is characterized by a single Higgs/confinement phase while the $\psi\eta$ model has two distinguishable phases with a phase transition
on the way. We will discuss this issue in Sec. \ref{sec:psieta}. 

Our results in the $\psi\eta$ and $\chi\eta$ models are essentially the same as 
obtained in \cite{Shrock,Poppitz}.
%except an interpretational difference and a relatively minor nuance.

\vspace{1mm} 

Our nomenclature is as follows (see also Fig. 1):\\

(i) $\psi\chi\eta$ model suggested in \cite{AS}. This model has $SU(N)_{\rm c}$ gauge symmetry and the fermion sector consisting of
the left-handed fermion matter fields  
\beq   
\psi^{\{ij\}}\;, \qquad  \chi_{[ij]}\;, \qquad    \eta_i^A\;,\qquad  (\small  {A=1,2,\ldots , 8})\,,
\eeq
where $\psi^{\{ij\}}$ is a two-index symmetric representation of $SU(N)_{\rm c}$,  $\chi_{[ij]}$ is an  anti-antisymmetric tensor,  and we have eight anti-fundamental fields $  \eta_i^A $. The index $i$ belongs to $SU(N)_{\rm c}$ while  $A$ to  the chiral  flavor $G_f=SU(8)_{\rm f}$ symmetry.  All spinor fields above (i.e. $\psi, \,\, \chi$ and $\eta$) are represented by undotted Weyl spinors.\\

(ii) $\chi\tilde{\eta}$ model. The gauge group is $SU(N)_{\rm c}$ while the fermion sector is composed of 
\beq    
 \chi_{[ij]}\,, \qquad     \tilde{\eta}^{i, \, A}\,,\qquad  (\small  {A=1,2,\ldots , N-4})\,. 
\eeq
The flavor symmetry of the model is $SU(N-4)_{\rm f}\times U(1)$.  \\

(iii)  $\psi\eta$ model. The gauge group is $SU(N)_{\rm c}$. The fermion sector includes
\beq    
 \psi^{\{ij\}}\,, \qquad     \eta_i^A\,,\qquad  (\small {A=1,2,\ldots , N+4})\,. 
\eeq
The flavor symmetry of the model is $SU(N+4)_{\rm f}\times U(1)$.   \\

The theory (i) on which we focus is asymptotically free, so that the interactions become strong in the infrared, but no gauge-invariant bifermion scalar condensates are possible.
It possess a nontrivial chiral symmetry $G_f = SU(8)$;  the requirement that  $G_f$ is either spontaneously broken (entailing the Nambu-Goldstone mesons) or realized linearly in the infrared, combined with the 't Hooft matching, is so strong that it allows us to determine the most likely dynamical $\chi$SB pattern. 

Even though gauge-singlet condensates of different types, such as four-fermion composites, $\sim \Psi^4$,  ${\bar \Psi}^2 \Psi^2$ or  bifermion condensates with gauge fields  (such as ${\bar \Psi} G \Psi$,  ${\bar \Psi} G G  \Psi$, etc.),  are in principle possible in the case at hand,  the 't Hooft matching in the linear realization cannot be achieved. As to the Nambu-Goldstone (NG) realizations, they appear to be highly contrived and, moreover,  the NG bosons couplings  to the appropriate flavor currents is hard to organize. 

  The other possibility of gauge-singlet composite massless fermions, which, for example, is long known to be a viable solution in the $\chi\tilde{\eta}$ and  $\psi\eta$  models \cite{Shrock,Poppitz}, is highly improbable for the  $\psi\chi\eta$ model. Saturating the flavor $G_f = SU(8)$ would in fact require order $N$ composite fermions all to remain massless,  and this is very unlikely in the absence of an $SU(N)$ symmetry in the IR.

In view of the above we explore the bifermion gauge-noninvariant condensates 
$ \sim \Psi^2$ in the hybrid $\psi\chi\eta$ model Higgsing
the gauge symmetry (more exactly, a part of it). One of the bi-fermion condensates is 
due to an operator in the fundamental representation, to which the
Fradkin-Shenker continuity applies. However, an important role belongs to another operator, in the adjoint representation, 
which leads to dynamical Abealization. 

 \begin{figure}[h]
\begin{center}
\includegraphics[scale=.8]{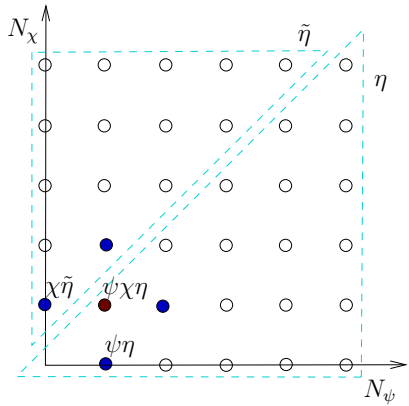}
\end{center}
\caption{\small A class of chiral QCD theories at large $N$ in the plane $(N_{\psi},N_{\chi})$. } 
\label{theories}
\end{figure}

 Figure \ref{theories} gives a schematic representation of the various  irreducibly chiral theories at large
  $N$ in which both $N_{\psi}$ and $N_{\chi}$ can go up to $5$ without loss of asymptotic freedom.  The number of fundamentals $\eta$ or anti-fundamental
$\tilde{\eta}$ is then fixed in order to cancel the gauge anomaly. The closed circles mark the  prototypes models disussed above.
Apart from a few exceptions where gauge invariant composite fermions can saturate the anomaly, as for  $(N_{\psi},N_{\chi}) = (1,0),(2,0),(0,1),(0,2)$, most theories of this class require, as the most plausible solution of the anomaly matching, the dynamical Higgsing by bifermion  gauge-noninvariant condensates.

One might suspect that the analysis of the above models (especially the $\psi\chi\eta$ model) phrased in terms of the gauge non-invariant condensates at strong coupling carries an ambiguity in contradistinction with, say, the Higgs mechanism at weak coupling. We want to emphasize that this is just a technically convenient language which in fact implies physically clear-cut and absolutely unambiguous predictions for measurable quantities in the infrared domain. The crucial prediction following from our consideration of the $\psi\chi\eta$ model is the existence in the physical spectrum of the theory $\sim N$ massless ``photons", in addition to an appropriate number of the massless fermions (having in essence the quantum numbers of quarks) which saturate the 't Hooft triangles. Thus our solution of the $\psi\chi\eta$ model is distinctly different from the previous considerations of chiral QCD based on the presumption that only two options are possible in the massless sector: either massless Nambu-Goldstone mesons or fine-tuned baryons. Our solution which goes under the code name of Dynamical Abelization predicts the emergence of the Coulomb phase (for $N>12$  the Coulomb phase is mixed with the ``confinement" phase). This is a qualitative difference and a new way to achieve the 't Hooft matching. Although it might be very difficult, it would be of paramount importance to check this result in lattice calculations.

The paper is organized as follows. Section \ref{psichi} is devoted to the $\psi\chi\eta$ model, the main subject of our studies.
 In Sec. \ref{wcsb} we argue that the suggested solution of the $\psi\chi\eta$ model, which, generally speaking, leads to Abelianization of the gauge group, is preferred while other solutions are unlikely.
In Secs. \ref{sec:chieta} and \ref{sec:psieta} we briefly comment on the $\chi\tilde\eta$ and $\psi\eta$ 
models emphasizing the Fradkin-Shenker continuity between two regimes of the joint confinement/Higgs phase. In Sec.~\ref{mac} the standard Maximal Attraction Channel (MAC)
arguments are summarized.

\vspace{1mm}

\section{\boldmath{$\psi\chi\eta$} model} 
\label{psichi}

The $\psi\chi\eta$ model is  asymptotically free, with the first coefficient of
the beta function 
\beq
  b=   \frac 13  \left( 9 N   -  8 \right)   \,.
\eeq
The global symmetry of the model is 
\be    G_f= SU(8) \times U_1(1)\times U_2(1)\times  {\mathbbm Z}_{N^*}\;.
\ee
where $U_{1,2}$ are anomaly-free combinations out of $U_{\psi}$, $U_{\chi}$, $U_{\eta}$, which can be taken e.g., as 
\beqn 
&&   U_1(1): \qquad   \psi\to  e^{i  \frac{\alpha}{N+2}} \psi\;; \qquad \eta \to   e^{- i \frac{\alpha}{8}} \eta\;;   
\nonumber \\[1mm]
&&   U_2(1): \qquad   \psi\to  e^{i  \frac{\beta}{N+2}} \psi\;; \qquad \chi \to   e^{- i \frac{\beta}{N-2}} \chi\;.  \label{nonanU1}
\eeqn
The discrete symmetry 
\be       {\mathbbm Z}_{N^*}\;, \qquad    N^* = {\rm GCD} \{N+2, N-2, 8\}  \label{vacua}
\ee
is a subgroup of  anomaly-free  discrete   ${\mathbbm Z}_{N+2}\otimes {\mathbbm Z}_{N-2} \otimes  {\mathbbm Z}_{8}$, which do not belong to 
$U_1(1)\times U_2(1)$. 

There are two fermion bilinear condensates,
\be
 \langle   \phi^{i A}  \rangle \sim  \langle  \psi^{ij}  \eta_j^A   \rangle \,,\quad A= 1,2, \ldots, 8\,,
\ee
and 
\be  \,\,\,  \langle   {\tilde \phi}^{i}_j \rangle = \langle  \psi^{ik} \chi_{kj}   \rangle\;,
\label{11}
   \ee
 which, we assume, play the crucial role in the dynamics.  The first one is in the fundamental representation of the gauge group, while the second is in the adjoint.
   
  \subsection{(Partial) Color-flavor locking} 
If 
\beq
\langle  \phi^{i A} \rangle = \brc  \psi^{ij}  \eta_j^A  \ckt  \propto  \Lambda^3 \delta^{iA} \neq 0\; ,\qquad i, A= 1,2, \ldots, 8\,,
 \label{eight}
 \eeq
the flavor $SU(8)$ is spontaneously broken, and so is the $SU(8)$ subgroup of $SU(N)_{\rm c}$. The global diagonal 
 $SU(8)$ (also known as color-flavor (cf) locked group) survives. The NG  excitations associated with the
 spontaneous breaking of $SU(8)_{\rm f}$ are eaten up by the gauge bosons of  $SU(8)_{\rm c}$; they Higgs this subgroup.
 Also unbroken is the $SU(N-8)_{\rm c}$ subgroup provided the second condensate, Eq. (\ref{11}),  does not develop.\footnote{ 
 It is known \cite{Fradkin} that if the order parameter is in the fundamental representation and can completely Higgs the $SU(N)_c$ gauge group there is no phase transition
 between the Higgs and confinement regimes; they both represent a single Higgs/confinement
 phase with a crossover somewhere between weak and strong coupling. 
 In this section this is irrelevant, but
 we will return to a more detailed discussion of this 
 point later.} 
 
Now we will argue that the system must develop also
the condensate in the adjoint representation of the form, 
\be   \langle  \tilde{\phi}^{i}_j \rangle \equiv  \langle  \psi^{ik} \chi_{kj}   \rangle = \Lambda^3   \left(\begin{array}{c|ccc|c}
 a \,  {\mathbf 1}_{8}   &  &  & &  \\ \hline  &  d_1  &  & & \\ &   & \ddots &  & \\ &  &  & d_{N-p}&
\\ \hline & &  &&   b  \,  {\mathbf 1}_{p-8}   \end{array}\right)\;,       \label{condensate}
\ee
where 
\be    8  a + \sum_i  d_i + (p-8)  b =0\;, 
\qquad     a, d_i, b \sim    O(1)\,, 
\ee
and  $d_i$'s are assumed to be all  distinct, and different either from $a$ or $b$. As we will see shortly, the integer parameter $p=12$ where we assume 
for the time being that $N \geq 12$. Smaller values of $N$ are discussed at the end of this subsection.

The symmetry breaking pattern is 
\beq
SU(N)_{\rm c} \times SU(8)_{\rm f}   \times U(1)^2  \to   SU(8)_{\rm cf} \times  U(1)^{N-p+1} \times SU(p-8)_{\rm c}\,.
\label{scenario1p}
\eeq
The $U(1)^2$ factor on the left-hand side of (\ref{scenario1p}) corresponds to the two non-anomalous $U_{1,2}(1)$ symmetries of (\ref{nonanU1}).
If the condensates (\ref{eight}) and (\ref{condensate})  are both nonvanishing, 
the anomaly-free  symmetries $U_{1,2}(1)$ are both spontaneously broken.\footnote{Unlike what happens in the reduced $\chi {\tilde \eta}$  model in Section \ref{sec:chieta},  here  the flavor $U_{1,2}(1)$ cannot mix with some Cartan subgroup of $SU(N)$ to form a new  unbroken    ``flavor''  $U(1)$ group(s).      }

The factor $U(1)^{N-p+1}$ on the right-hand side of
 Eq.~(\ref{scenario1p}) represents  the unbroken Cartan subgroups of $SU({N-p})$ times two more  $SU(N)$ Cartan subalgebras,  which can be taken, e.g.,  as  
\be   
 \left(\begin{array}{ccc}{\mathbf 0}_8 &  &   \\   &   - \tfrac{1}{N-p}  {\mathbf 1}_{N-p}  &   \\  &   &  \tfrac{1}{p-8} {\mathbf 1}_{p-8}\end{array}\right)\;,       \quad  
  \left(\begin{array}{ccc}    \tfrac{1}{8}  {\mathbf 1}_8 &  &   \\   &    {\mathbf 0}_{N-p} &   \\  &   &  -  \tfrac{1}{p-8} {\mathbf 1}_{p-8}\end{array}\right)\;.    
    \ee 

Let us  denote the indices from $SU(p-8)$ by bars.
Then all gluons with unbarred indices 
(except those from the Cartan subalgebra of $SU(N-8)$) get masses. We have no confinement for unbarred indices.   It is only $SU(p-8)$ that truly confines.
Most of the fermions are unconfined. We could reach the triangle saturation \`a la 't Hooft as follows. 

Let us consider ``baryons" --   massless (by assumption)  fermion states which are constructed from condensates and the $\eta$ fields,
\beq
      B^{\{AB \}}=\psi^{\{ij\}} \eta_j^A \eta_i^B\Big|_{A,B \,\, {\rm symm} }  \sim     \langle \phi^{i A} \rangle \eta_i^B\Big|_{A,B \,\, {\rm symm} }    \eeq
transforming in the symmetric representation of 
$SU(8)_{\rm cf}$, which can be achieved by an appropriate spin index convolution, and 
\beq      
{\tilde B}^A_j=\psi^{ik} \chi_{kj} \eta_i^A   \sim   \langle {\tilde \phi}^{i}_j\rangle   \eta_i^A   \, .
\label{baryonp}
\eeq
The former is a symmetric tensor of $SU(8)_{\rm cf}$.  It contributes  $8+4=12$ to the $SU(8)^3$ triangle. 
The baryons ${\tilde B}^A_j$ are in the  fundamental representation of both  color and  flavor groups.
 In the scenario of partial Higgsing of the gauge group  (\ref{condensate}), taking $j=9,10,\dots N-p+8$, 
we find that ${\tilde B}^A_j$ contributes 
$    N-p  $ to the $SU(8)^3$ triangle anomaly. We observe that
 the sum of  the contributions from  $   B^{\{AB \}}$ and  $ {\tilde B}^A_j$   is 
\beq    
12+    N-p  \,,
\label{fourteen}
\eeq 
which perfectly saturates the 't Hooft anomaly matching for $SU(8)^3$ if and only if $p=12$.

The symmetry breaking pattern (\ref{scenario1p})  is, therefore,
\beq
SU(N)_{\rm c} \times SU(8)_{\rm f}   \times U(1)^2  \to   SU(8)_{\rm cf} \times  U(1)^{N-11} \times SU(4)_{\rm c}\,.
\label{scenario1}
\eeq
It is easy to see that the overall rank of the gauge factors on the right-hand side is $N-1$, i.e. exactly the same as in the left-hand side.
Since the anomaly-free global $U(1)^2$ symmetries on the left-hand side of (\ref{scenario1}) are both broken by  the condensates 
(\ref{eight}) and (\ref{condensate}) we need not 
worry about the $U(1)$-$SU(8)^2$  or   $U(1)^3$  triangles.   There must exist two NG  bosons associated with these non-anomalous $U(1)$'s,   which are presumably the phases of the condensates (\ref{eight}) and (\ref{condensate}).     The condensates take the form
\beq
 \brc  \phi^{i A}   \ckt  = \Lambda^3 \left(\begin{array}{c} c  {\mathbf 1}_{8}  \\ \hline
  \\
{\mathbf 0}_{N-8,8}\\
\\
\end{array}\right)\;, \qquad 
 \langle  \tilde{\phi}^{i}_j \rangle  = \Lambda^3   \left(\begin{array}{c|ccc|c}
 a \,  {\mathbf 1}_{8}   &  &  & &  \\ \hline  &  d_1  &  & & \\ &   & \ddots &  & \\ &  &  & d_{N-12}&
\\ \hline & &  &&   b  \,  {\mathbf 1}_{4}   \end{array}\right)\;,   \label{conden}
\eeq
where 
\be    8 a + \sum_{i=1}^{N-12}  d_i + 4  b =0\;, 
\qquad     a, d_i, b \sim    O(1)\;. 
\ee
 The theory dynamically Abelianizes (in part). $SU(8) \subset SU(N)$ is completely Higgsed but due to color-flavor (partial) locking no NG bosons appear  in this sector (the would-be NG bosons make the $SU(8) \subset SU(N)$  gauge bosons massive.) Only $SU(4) \subset SU(N)$ remains unbroken and confining. 
The remainder of the gauge group Abelianizes.   The baryons ${\tilde B}^A_j \sim  \eta_j^A$  ($9 \le j \le N-4$) and     $B^{\{AB \}}$, symmetric in the flavor indices,\footnote{If the massless  $B^{\{AB \}}$ were antisymmetric in $(A\leftrightarrow B)$, they would contribute $8-4=4$ to the $SU(8)$ anomaly.  We would then need $N-4$  massless fermions of the form  ${\tilde B}^A_j \sim  \eta_j^A$,
but this is impossible as the latter arise from the Abelianization of the rest of the color gauge group,  $SU(N-8)$. }   remain massless  and together saturate the  't Hooft  anomaly matching condition for $SU(8)$.  

Actually,   it is possible that the color-flavor locking occurs with  split $SU(8)_{\rm cf}$. If  $\langle  \phi^{i A} \rangle $ 
takes the form,
    \beq
 \brc  \phi^{i A}   \ckt  = \left(\begin{array}{ccc}c_1 {\mathbf 1}_{A_1} &    &  \\  &   \ddots &  \\ &    & c_h {\mathbf 1}_{A_h} \\ \hline
&{\mathbf 0}_{N-8,8}&\end{array}\right)\;, \qquad   \sum A_i=  8\;,  \qquad 
\eeq
and $\langle  \tilde{\phi}^{i}_j \rangle$ is the same as before, 
the pattern  (\ref{scenario1})   is replaced by 
\beq
SU(N)_{\rm c} \times SU(8)_{\rm f}   \times U(1)^2  \to   \prod_i SU(A_i)_{\rm cf} \times  \prod U(1)_{\rm cf}  \times    U(1)^{N-11
} \times SU(4)_{\rm c}\,.
\label{scenario1bis}
\eeq
All various triangle anomalies associated with $ \prod_i SU(A_i)_{\rm cf} \times  \prod U(1)_{\rm cf}$  are seen to be fully saturated by the massless baryons $ B^{\{AB \}}$
and weakly coupled  ${\tilde B}^A_j$'s.\footnote{As miraculous as it might look, this is simply due to the fact that  $ \prod_i SU(A_i)_{\rm cf} \times  \prod U(1)_{\rm cf}   \subset  SU(8)_{cf}$. }        In this case, there are massless NG bosons associated with  symmetry breaking, 
\be      \frac{SU(8)_{cf}} { \prod_i SU(A_i)_{\rm cf} \times  \prod U(1)_{\rm cf}  }\;.
\ee

Thus we found possible dynamical scenarios for our system, (\ref{scenario1}) or  (\ref{scenario1bis}).  They differ in details of color-flavor locked symmetries, but both hinge upon a partial color-flavor diagonal symmetry and a (partial) dynamical Abelianization of the gauge symmetry. They collaborate to realize correctly the anomalies associated with the original flavor symmetry $SU(8)$.   

\vspace{2mm}

Clearly, however,  these solutions are   possible only for $N \ge 12$.    For smaller $N$  we might resort to the 
the idea that  the color-flavor locked global symmetry  occurs in a smaller block.
It is tempting then to consider the symmetry breaking pattern 
\beqn   
&&  SU(N)_{\rm c}\times SU(8)_{\rm f} \times U(1)^2  \nonumber \\[2mm]
&\to&    SU(M)_{\rm cf} \times  U(1)^{N-p+1} \times SU(p-M)_{\rm c}  \times SU(8-M)_{\rm f} \times U(1)_{\rm cf}
\label{scenariopp}
\eeqn
with $M < 8$  (and, of course,  $M \leq N$), in other words,  to assume the $\psi \eta$  condensate of the form,
\beq
\langle  \phi^{i A} \rangle = \brc  \psi^{ij}  \eta_j^A  \ckt  =   \Lambda^3 \delta^{iA} \neq 0\,,  \qquad i, A= 1,2, \ldots, M\,, \quad  M< 8\, .
 \label{M}
 \eeq
Actually, it can be easily checked that the fermions  $ B^{\{AB \}}$ and  ${\tilde B}^A_j$ cannot saturate the anomalies $SU(M)_{\rm cf}^3 $,  $ SU(8-M)_{\rm f}^3$, and other triangles involving $U(1)_{cf}$,
for any choice of $p$.  Hence the dynamical scenario presented in  (\ref{scenariopp}),(\ref{M}), should be excluded. 

One is left with the possibility of full Abelianization as the only dynamical scenario for $N<12$. 

\subsection{Full Abelianization}
\label{fulla}

Another option, which so far is the only viable one left  for $N$ smaller than $12$, is that no color-flavor locking takes place
($\langle  \phi^{i A} \rangle=0$).
The flavor symmetry remains unbroken:
  \beq     
  SU(N)_{\rm c} \times SU(8)_{\rm f} \times U(1)^2 \to U(1)^{N-1} \times  SU(8)_{\rm f}\, .
  \label{scenario3}
  \eeq
The gauge group dynamically Abelianizes completely.  
The fields  $\eta^A_i$ are all massless and  weakly coupled (only to the gauge bosons from the Cartan subalgebra which we will refer to as the photons; they are infrared free) in the infrared, together with the $N-1$ photons. The anomaly matching is trivial. 
 
\subsection{Remarks}

We are unable, for general large  $N$,  to  decide which of the dynamical scenarios  (\ref{scenario1}), (\ref{scenario1bis}) or (\ref{scenario3})  is actually  realized. They represent a few  
possible dynamical scenarios, with color-flavor locking and (partial) dynamical Abelianization,  which produce
 some massless baryon-type composites fermions, together with weakly coupled  $\eta$ fields associated with the Abelianization. 
It is interesting that in some other chiral theories (discussed in  Sections \ref{sec:chieta} and \ref{sec:psieta}), there is no Abelianization. 

 Abelianization is a ubiquitous phenomenon in ${\cal N}=2$ supersymmetric gauge theories, where scalar fields in the adjoint representation are present in the UV, and 
 the perturbative potential has flat directions so that   $\brc \phi_{\rm adjoint}\ckt \ne 0$  at a general point on the vacuum moduli space.   Even though appealing, however,  it seems to be 
highly unlikely that the Seiberg-Witten duality\cite{SW}  (the low-enegy degrees of freedom are the magnetic monopoles and dyons, the associated  gauge groups being dual, magnetic 
 $U(1)^{N-1}$'s)  can be realized dynamically in our model, in view  of absence of the flat directions. 
The strength of the effective adjoint scalar condensate in our case is of  the order of  $\Lambda^3$. The  associated fermions 
$\eta_i^A$ now interact only with 
the $U(1)$ fields, and thus  become weakly coupled in the infrared.  Somewhat analogous phenomena in  ${\cal N}=2$ supersymmetric gauge theories would be the 
rather trivial  cases of quark singularities at large values of adjoint scalar vacuum expectation values (VEV's). 

\subsection{Planar equivalence}

The $\psi\chi\eta$ model was argued \cite{AS} to be planar equivalent to ${\cal N}=1$ super-Yang-Mills theory (SYM).
The arguments were based on the assumption that both theories are realized in the confining regime.
Since, as we see,  the $\psi\chi\eta$ model is realized in the Higgs regime (full or partial Higgsing) the planar equivalence fails. 
One can present an alternative consideration demonstrating the failure of the planar equivalence between SYM and the
Armoni-Shifman model \cite{AS}.  This consideration follows the line of reasoning of \cite{unsal}.

If we analyze both theories on a cylinder $R_3\times S_1$ with $r(S_1) \ll \Lambda^{-1}$ we will discover that 
for SYM theory the ${\mathbbm Z}_{N}$ center symmetry of $SU(N)_c$ is preserved, implying a smooth transition to 
$r(S_1) \gg \Lambda^{-1}$ and confinement in the $R_4$ limit. The SYM theory is $C$ symmetric, and $C$ invariance is not spontaneously broken \cite{unsal,aaunsal}. 

At the same time, the Armoni-Shifman model \cite{AS} explicitly breaks $P$ and $C$ invariances, its Lagrangian preserves only $CP$. 
Moreover, its vacuum structure is completely different from that of SYM and is characterized by broken 
${\mathbbm Z}_{N}$ center symmetry of $SU(N)_c$  \cite{CSU}. The statement of $N$ isolated vacuum states in SYM 
is also not inherited by the $\psi\chi\eta$ model because of  the NG boson (see also  (\ref{vacua})).

\section{Why chiral symmetry breaking patterns without Higgsing are disfavored}
\label{wcsb}

Let us return to the $\psi\chi\eta$ model discussed in Sec. \ref{psichi} and analyze the chiral $SU(8)$ symmetry under the assumption that the gauge $SU(N)$ symmetry is not Higgsed, completely or partially, and the theory is realized in the confinement regime similar to that of QCD. The simplest option is that no fermion condensates are formed and $SU(8)_{\rm f}$ remains unbroken. Then we will have to match $SU(8)_{\rm f}^3$ triangle and two $SU(8)_{\rm f}^2\times U(1)$ triangles.  To find a set of massless gauge-invariant composite fermions (baryons), able to reproduce the  correct
UV    $SU(8)_{\rm f}^3$ and  $SU(8)_{\rm f}^2\times U(1)$ anomalies  ($= N$)  appears to be a hopeless task. 

% It is is not difficult to establish that for generic values of $N$ that the UV coefficients in the triangles ($N$)  can not be matched to those obtained  by color-singlet baryons in the infrared. The 't Hooft matching fails. 

Now let us analyze color-singlet condensates. The simplest relevant one is the four-fermion
condensate 
\be   {\overline  {\psi^{ik}  \eta_k^A }}  \,   \psi^{ij}  \eta_j^B     \,.
\label{bks67}
\ee
Related NG excitations must  be (derivatively) coupled to the current
\beq
\left(j_{\alpha\dot\alpha}\right)^B_A = \bar\eta_{\dot\alpha\, A}\eta_\alpha^B\,. 
\eeq
Note, however, that the condensate (\ref{bks67}) is purely real and therefore
does not break two anomaly-free $U(1)$ chiral symmetries.  The condensate  (\ref{bks67}) if existed would break
$SU(8)_{\rm f} \to U(1)^7$. These $U(1)$ factors are the Cartan subalgebra of $SU(8)_{\rm f} $, not to be confused with
with the  anomaly-free $U(1)$ chiral symmetries of the type $U(1)_\eta + C_1 U(1)_\psi$ and $U(1)_\eta + C_2 U(1)_\chi$ (here $C_{1,2} $ are numerical constants). 

To break the flavor symmetry completely, we must add another condensates, e.g.
\be   \Theta^{CD}_{AB} =    {\overline {( \eta_{[i}^A \eta_{j]}^B )} }    (\eta_{[i}^C \eta_{j]}^D)\qquad  \mbox{or} \qquad 
   \Gamma^B_A=     {\overline { (\chi_{[ij]} \eta_k^A) } }    (\chi_{[ij]} \eta_k^B) \,.
\ee
However, the symmetries $U(1)_\eta + C_1 U(1)_\psi$ and $U(1)_\eta + C_2 U(1)_\chi$  mentioned above remain unbroken. This means that the triangle diagrams due to these symmetries must be matched \`a la 't Hooft by saturating them by massless baryons, for instance, $\psi \eta \eta$.  Again, the matching does not seem possible at arbitrary $N$. 

Finally, a chiral four-fermion condensate  
\be     \psi^{ij}   \chi_{[jk]}   \,  \psi^{k \ell}   \chi_{[\ell i]} \; \label{chiral}
\ee
 could  break  both nonanomalous $U(1)$ symmetries,  but not  $SU(8)$:  it does not 
affect the possible mechanism of realization of the chiral $SU(8)$ symmetries at low energies. 

All in all, the dynamical Higgs mechanism by formation of gauge-dependent bifermion condensates discussed in the previous sections appears to be 
the most natural way of realizing the chiral flavor symmetries at low energies, in the class of models considered in the present paper. 

\section{\boldmath{$\chi\tilde\eta$} model  \label{sec:chieta}}  

In this and the subsequent section (Sec. \ref{sec:psieta}) we
 will discuss the $\chi\tilde\eta$ and $\psi\eta$ models.  
 These two models were thoroughly studied in \cite{Shrock,Poppitz} with the conclusion that the chiral symmetry 
is unbroken, and the 't Hooft triangles are saturated by massless composite baryons. In view of our previous remarks on 
the Fradkin-Shenker continuity the solution of \cite{Shrock,Poppitz} in the $\chi\tilde\eta$ case is physically indistinguishable from ours since this model 
 belongs to the joint confinement/Higgs phase with no phase transition between the two regimes. (See a reservation in the $\psi\tilde\eta$ case in Sec. 
 \ref{sec:psieta}.)
 
Let us consider the $SU(N)$ gauge theory with the  chiral fermion sector\,\footnote{The $SU(5)$ model with one quintet and one ${\overline{10}}$ was used for Grand Unification. }
\beq
     \chi_{[ij]}\,, \quad     \tilde{\eta}^{i\, A}\,, \qquad  A=1,2,\ldots , N-4\,. 
\eeq

Compared to the $\psi\chi\eta$ model we dropped the field $\psi$ and adjusted the number of $\eta$'s (taken in the fundamental, rather than antifundamental representation of $SU(N)_c$)
to keep the theory free of the gauge anomaly. 
The global symmetry of  this model is  $SU(N-4)_{\rm f} \times { U}(1)$. The latter $U(1)$ is an anomaly-free linear combination of two
$U(1)$ symmetries of the fermion sector. The second $U(1)$ is anomalous.
The $\chi\tilde{\eta}$ model is known to be asymptotically free (AF), with the first coefficient of
the beta function 
\beq
  b=   \frac 13\left[11 N   -  (N-2)  -  (N-4)\right] =  3 N +2 \,.
\eeq
The theory becomes strongly coupled in the infrared. 

Having no  {\em adjoint} complex scalar order parameter bilinear in the fermion fields  (as $\psi \chi $ in the $\psi\chi\eta$ model considered earlier), we do not expect dynamical Abelianization to occur.  
Let us assume that the scalar 
\beq
    {\varphi}_j^A     \equiv   \chi_{[ij]} \,  \tilde{\eta}^{i\, A}
    \label{f25}
\eeq
acquires a  color-flavor locked  diagonal  VEV,\footnote{This option was mentioned as one of a number of possibilities in \cite{unsalms}.} as in (\ref{eight}),
\beq   
 \langle   {\varphi}_j^A  \rangle = \, c \,  \Lambda^3   \delta_j^A\neq 0\,, \qquad  j, \, A =1,2, \ldots ,  N-4\;.   
 \label{condensechiq}
\eeq
Then the gauge and global symmetry breaking pattern  is
\beq
    SU(N)_{\rm c}  \times  SU(N-4)_{\rm f} \times U(1)  \to    SU(N-4)_{\rm cf}  \times U(1)^{\prime}   \times   SU(4)_{\rm c}\,.
    \label{b27}
\eeq
where $ SU(N-4)_{\rm cf}$ is a color-flavor diagonal symmetry, and $U(1)^{\prime}$ is a linear combination of the anomaly-free global $U(1)$ and a $U(1)$ subgroup of $SU(N)_{\rm c}$,  namely, 
$$   \left(\begin{array}{cc}\tfrac{1}{N-4} {\mathbf 1}_{N-4} & 0 \\0 & -  \tfrac{1}{4} {\mathbf 1}_4\end{array}\right).
$$  

The ranks of the gauge groups in both sides of Eq. (\ref{b27}) match.     The $SU(4)_{\rm c}$ factor on the right-hand side of (\ref{b27}) confines. 

%\marginpar{\tiny     The rank of $SU(N)_c$ on lhs is $N-1$; on the r.h.s., rank of $SU(N-4)_{cf}$ is  $N-5$, that of $SU(4)_c$ is 3
%and the rank of $U(1)^{\prime}_{cf}$ is 1. Altogether we have 
%$N-1$, right? }

The broken gauge fields of
\be     SU(N)_{\rm c} / SU(4)_{\rm c} 
\ee
become all massive by the standard Higgs mechanism;  the gauge bosons belonging to $SU(N-4)_{\rm c} \subset SU(N)_{\rm c} $ remain degenerate, reflecting the 
color-flavor locked global $SU(N-4)_{\rm cf} $ symmetry.  There are no massless Nambu-Goldstone bosons.    

The massless baryons similar to (\ref{baryonp}) take the form
\beq
 \tilde{B}^{\{AB\}} = \chi_{[ij]}\, \tilde{\eta}^{i\, A}\, \tilde{\eta}^{j\, B}  \sim    \langle {\varphi}_j^A  \rangle\,  \tilde{\eta}^{j\, B} 
 \label{ss32}
\eeq
transforming in the symmetric representation of 
$SU(N-4)_{\rm cf}$.
Then the  $\tilde{B}^{\{AB\}}$  multiplet  contributes to  the $ SU(N-4)_{\rm cf}^3$ triangle as
\beq
   N-4  + 4 = N \,.
\eeq
This exactly matches the ultraviolet anomaly in the corresponding flavor symmetry triangle in the $\chi\eta$ model, which reduces to 
$N$  from the original quarks   $\tilde{\eta}^{i\, A}$  which belong to the fundamental representation of the UV gauge symmetry. 

This system was studied earlier  \cite{Shrock,Poppitz},  in particular by Appelquist, Duan and Sannino  \cite{Shrock},  where the massless baryons $ \tilde{B}^{\{AB\}}$ were shown to saturate the triangles with respect to the  full unbroken global symmetries  $SU(N-4)_{\rm  f}  \times U(1)$. 
(This was pointed out to us also by E. Poppitz). 
%In the above publications the parameter $c$ in Eq.~(\ref{condensechiq})) was set to zero and the anomaly matching condition 
%with respect to the full global symmetry group was shown to be satisfied by  $ \tilde{B}^{\{AB\}}$.   
In the above publications the possibility  $c\ne 0$ in Eq.~(\ref{condensechiq}) was also analyzed, and the anomaly matching condition 
with respect to $SU(N-4)_{\rm cf}  \times U(1)^{\prime} $  was shown to be satisfied by the same set of fermions $ \tilde{B}^{\{AB\}}$.  

The agreement between the two distinct descriptions 
is just another demonstration of the Fradkin-Shenker continuity.

\section{\boldmath{$\psi\eta$} model \label{sec:psieta}  } 

Consider  another reduction of the $\psi \chi\eta$ model. Namely, we drop the field $\chi$ and adjust
 the number of $\eta$'s
to keep the theory free of the gauge anomaly. For the 
 $SU(N)$ gauge group the fermion sector is 
\beq
   \psi^{\{ij\}}\,, \quad    \eta_i^B\, , \qquad  B=1,2,\ldots , N+4\,. 
\eeq
The symmetry of this model is 
\beq
SU(N)_{\rm c} \times SU(N+4)_{\rm f} \times {U}(1)\,,
    \eeq
   where ${ U}(1)$ is an anomaly free-combination of  $U(1)_{\psi}$ and  $U(1)_{\eta}$.
This model is asymptotically free as the first coefficient of the $\beta$ function is
 \beq    
 b= \frac 13\left[11 N   -  (N+ 2)  -  (N+4) \right] =  3 N  -  2 \;.
\eeq
The theory is strongly coupled in the infrared. 
Again, we do not have a dynamical bifermion scalar operator in the adjoint representation of the gauge group.
As previously (see (\ref{f25})), we 
assume that the condensate of $\Phi^{j\,B}$ develops,
\beq  
\langle \Phi^{j\,B}  \rangle  =
\langle  \psi^{ij}   \eta_i^B  \rangle  = c \Lambda^3 \delta^{j B} \neq 0\,,\qquad j,\, B = 1,2,..., N\,.  
\label{f33}
\eeq
Then the symmetry breaking pattern takes the form   
\beq   
SU(N)_{\rm c} \times SU(N+4)_{\rm f} \times  U(1)  \to  SU(N)_{\rm cf} \times  U(1)^{\prime}  \times  SU(4)_{\rm f}\;.     \label{flavors}
\eeq
The non-anomalous $U(1)$, a linear combination of  $U(1)_{\psi}$ and  $U(1)_{\eta}$
is spontaneously broken by the condensate (\ref{f33}), but a linear combination with a $U(1)$ subgroup of $ SU(N+4)_{\rm f}$,  
$$   
\left(\begin{array}{cc} \tfrac{1}{N} {\mathbf 1}_{N} & 0 \\0 & -  \tfrac{1}{4} {\mathbf 1}_4\end{array}\right),
$$
which we call $U(1)^{\prime} $, remains unbroken.  

There are 
\be   (N+4)^4-1 -  \{  N^2-1 +  15\} =    8N +1
\ee
physical massless NG  bosons in this system.  $SU(N)_{\rm c} $ gauge bosons become all massive by the Higgs mechanism, maintaining 
their mass degeneracy due to the color-flavor locking. 

Since we have $SU(N)_{cf}$ and $SU(4)_f$ as unbroken global symmetries all NG bosons must belong to
certain representations of the above groups. It is easy to see that there two bi-fundamental bosons with regards to  $SU(N)_{cf}\times SU(4)_f$,
which constitute $2\times N \times 4$ NG bosons. The remaining NG boson  $\sim {\rm Im}  \sum_{j=1}^N \Phi^{jj} $ is a singlet.

Now, let us check that  the $SU(4)_{\rm flavor}^3$  and  $SU(N)_{\rm cf}^3$  triangles  are  saturated by massless baryons
\beq   
B^{[AB]} =  \langle\phi^{j\,A}\rangle \eta_j^B=\langle  \psi^{ij}  \eta_i^A \rangle  \eta_j^B\;,   \qquad   A,B = 1,2,\ldots, N 
\label{ss39}
\eeq
and 
\beq  
\tilde{B}^{[AB]} =  \langle\phi^{j\,A}\rangle \eta_j^B=\langle  \psi^{ij}  \eta_i^A \rangle  \eta_j^B\,,   \quad   A = 1,2,\ldots, N\,,  \quad   B=N+1, \ldots N+4\,.  
\label{ss40}
\eeq
Here we must choose 
$B^{AB}$  in the antisymmetric representation of  the  $SU(N)_{\rm cf}$ group while  ${\tilde B}^{A B} $ is in the $({\underline N}, {\underline 4})$ of  $SU(N)_{\rm cf} \times SU(4)_{\rm flavor}$.

The $SU(N)_{\rm cf}^3$ triangle is indeed saturated in the infrared by the contributions of $B^{AB}$ and  ${\tilde B}^{A B}$:
\beq    N-4  +   4  = N\;.
\eeq 
The $SU(4)_{\rm f}^3$ triangle is saturated by   ${\tilde B}^{A B}$ alone  which produces the coefficient   $N$.  

This model was also analyzed in \cite{Shrock,Poppitz}, where massless composite baryons
corresponding to the interpolating operators $ \psi^{ij}  \eta_i^A   \eta_j^B$ (Eqs. (\ref{ss39}) and (\ref{ss40}))
were shown to saturate the  anomalies both with respect to $SU(N)_{cf}\times SU(4)_f$ and to the triangles involving $U(1)^{\prime}$. 
The possibility of $c=0$ in (\ref{f33}) was considered and  
it was shown that the same composite fermions $B^{[AB]}$ satisfied the anomaly matching in $SU(N+4)_{\rm f} \times  U(1)$. 
Both in the $\chi {\tilde \eta}$ model of Section \ref{sec:chieta} and in the $\psi \eta$ model of this section the same massless composite baryons 
are found to saturate the anomaly  matching conditions, whether or not the global symmetry is partially broken dynamically by condensates (\ref{condensechiq})
or (\ref{f33}).  (See in particular, Shi and Shrock \cite{Shrock}). 

There is a drastic distinction in the application of the Fradkin-Shenker continuity in these two cases, however. Indeed, in the $\chi\eta$ model 
the hadron spectrum is smooth in the sense that there is no phase transition between the ``Higgs regime'' and ``confining regime". In the $\psi\eta$ model the situation is different. The saturation of the 't Hooft triangles does not distinguish between the two regimes. The spectrum analysis does! In the first case (the Higgs phase)
we have the NG bosons in the spectrum, while in the second (confining) case no massless bosons appear. Thus, there is a phase transition.
At the moment, we do not know to which side the $\psi\eta$ model belongs. 

We conclude by noting that the Fradkin-Shenker continuity fails if the flavor chiral symmetry has a larger rank than the gauge symmetry (assuming that
that all matter fields are in the fundamental representation of the both groups).

\section{Extended \boldmath{$\psi\chi\eta$} models   \label{sec:expsichieta}  }

In the examples below the number of the fields $\eta$ in the fundamental representation of the gauge
group is established from the condition of the absence of the gauge anomalies.

Let us start from the $SU(N)$ gauge model with the chiral fermion sector
\beq
   \psi^{\{ij\},\, A}\;, \qquad \chi_{[ij]} \;, \qquad  \eta^B_j\;,  \qquad    A=1,2,\,\quad  B=1,2,\ldots, N + 12\,.
\eeq
The symmetries of the theory are
\beq
SU(N)_{\rm c}  \times SU(2)_{{\rm f}} \times SU(12+N)_{{\rm f}} \times U(1)^2\;. 
\eeq
The first coefficient of the $\beta$ function is
\beq  
b= \frac 13\left[ 11 \,N - 2 (N+2)-  (N-2)  -  (12+N) \right] = \frac 13\left(  7N -14 \right)\,.
\eeq
The triangle anomalies to be matched are $SU(12+N)_{{\rm f}}^3$\,. (Note that 
$SU(2)$ has no triangle anomalies.)

Assuming that 
\beq  \langle  \psi^{\{ij\},\, 1} \chi_{[ik]}  
\rangle = C^j \, \Lambda^3\, \delta^j_k \neq 0\,, \qquad j,k =1, 2, \ldots , N
\eeq
we break $SU(N)_{\rm c}$ down to the Cartan subgroup.  If also 
\be   \langle  \psi^{\{ij\},\, 2} \chi_{[ik]}  
\rangle  \neq 0\,, \quad j,k =1, 2, \ldots , N \;,  \ee
they can  together Higgs gauge group completely.   If   $  \langle  \psi^{\{ij\},\, 2} \chi_{[ik]}  
\rangle =  0 $,   instead,  the Cartan subgroup of $SU(N)_{\rm c}$  survives in the infrared.  
The fermion fields  $\eta^B_j$  are unconfined in the former case, or 
are  weakly coupled to the $U(1)^{N-1}$ in the latter  case. In either case,  the matching of the  $SU(12+N)_{\rm f}$
't Hooft triangles is trivial. 

%The $SU(N+12)$ triangles gives $N$ in the UV.   The condensates
%\beq  \langle  \psi^{\{ij\},\, A} \chi_{[ij]}  \rangle\;, \qquad  \langle  \psi^{\{ij\},\, A} \eta^B_j \rangle
%\eeq

The second extended $\psi\chi\eta$ model 
with the chiral fermion sector
\beq   
\psi^{\{ij\}}\,, \quad \chi_{[ij]}^A \,, \quad  \tilde{\eta}^{B\,j}\,, \quad    A=1,2,\,\quad  B=1,2,\ldots, N-12\,,
\eeq
can be dealt with exactly in the same way as above.

\section{Additional support from the Maximal Attractive Channel (MAC) argument}
\label{mac}

A general idea that color-nonsinglet bifermion condensates may form and induce the dynamical gauge symmetry breaking  was proposed long ago in  Raby et. al.\cite{Raby} basing on the so-called MAC argument. 
Even though we do not rely on the MAC criterion to decide what happens in our chiral models, and do not 
follow the rules  proposed in  \cite{Raby}, it is nonetheless suggestive to compare the strength of one-gluon 
 exchange force in various bifermion scalar  channels, formed by two out of the three types of fermions, $\psi$, $\chi$ and $\eta$. Some of the most probable
 channels are  
 \beqn & & A:  \qquad  \psi \left(\raisebox{-2pt}{\yng(2)}\right) \, \psi \left(\raisebox{-2pt}{\yng(2)}\right)   \quad  {\rm forming}  \quad   \,  \raisebox{-6pt}{\yng(2,2)}\;;
\nonumber \\  & & B:  \qquad  \chi \left(\bar{\raisebox{-9pt}{\yng(1,1)}}\right) \, \chi  \left(\bar{\raisebox{-9pt}{\yng(1,1)}}\right)  \qquad \ \   {\rm forming}  \quad \bar{\raisebox{-12pt}{\yng(1,1,1,1)}}\;;
\nonumber\\  &&C:  \qquad  \eta \left(\bar{\raisebox{-2pt}{\yng(1)}}\right) \, \eta \left(\bar{\raisebox{-2pt}{\yng(1)}}\right)   \quad  \qquad \ \  {\rm forming}  \quad    \bar{\raisebox{-9pt}{\yng(1,1)}} \;;
\nonumber\\ && D:  \qquad  \chi  \left(\bar{\raisebox{-9pt}{\yng(1,1)}}\right) \, \eta\left(\bar{\raisebox{-2pt}{\yng(1)}}\right)   \qquad  \  \  {\rm forming}  \quad   \bar{\raisebox{-9pt}{\yng(1,1,1)}}\;;
\nonumber \\  && E:  \qquad   \psi  \left(\raisebox{-2pt}{\yng(2)}\right)   \, \chi \left(\bar{\raisebox{-9pt}{\yng(1,1)}}\right)  \quad \  \ {\rm forming ~ an ~ adjoint ~ representation}\,\, ({\tilde \phi});\ 
\nonumber\\ && F:  \qquad \psi  \left(\raisebox{-2pt}{\yng(2)}\right)   \, \eta \left(\bar{\raisebox{-2pt}{\yng(1)}}\right)  \quad   \quad\  {\rm forming}  \quad   \raisebox{-2pt}{\yng(1)}\; \,\, \,  ({\phi});
 \eeqn
 The one-gluon exchange strength turns out to be, in the six cases above, proportional to  
 \beqn  & &  A:  \qquad     \frac{2 (N^2-4)}{N} -     \frac{ (N+2)(N-1)}{N}  -     \frac{ (N+2)(N-1)}{N}    = - \frac{2 (N+2)}{N}  \;;
\nonumber \\ & &  B:  \qquad     \frac{2 (N+1)(N-4))}{N} -     \frac{ (N+1)(N-2)}{N}  -     \frac{ (N+1)(N-2)}{N}    = - \frac{4 (N+1)}{N}  \;;
\nonumber\\   & &C:  \qquad     \frac{(N+1)(N-2))}{N} -     \frac{ N^2-1}{2N}  -     \frac{ N^2-1}{2N}    = - \frac{ N+1}{N}  \;;
\nonumber\\  & & D:  \qquad     \frac{3 (N+1)(N-3))}{2N} -     \frac{ N^2-1}{2N}  -     \frac{ (N+1)(N-2)}{N}    = - \frac{2N+2}{N}  \;;
\nonumber\\  & &  E:  \qquad   N-     \frac{ (N+2)(N-1)}{N}  -     \frac{ (N+1)(N-2)}{N}    =   - \frac{N^2-4}{N}  \;;
\nonumber \\ & &  F:  \qquad     \frac{N^2-1}{2N} -     \frac{N^2-1}{2 N}  -     \frac{ (N+2)(N-1)}{N}    = - \frac{(N+2)(N-1)}{N}  \;,
 \eeqn
 respectively.
 We note that the      ${\tilde \phi}$   ($\psi  \chi$)    and  $\phi$   ($\psi \eta$) condensates considered by us   (cases E and F, respectively)     correspond  precisely to the two most attractive channels, at large $N$:  
 their attraction strength  scales as  $O(N)$ in contrast to the other four channels which scale as   $O(1)$.

\section*{Acknowledgments}

We would like to thank R. Shrock and E. Poppitz who drew our attention to \cite{Shrock,Poppitz} and a number of related publications.
Their input was very important.

The work of M.S. was supported in part by DOE grant DE-SC 0011842 and  by the National Science Foundation under Grant No. NSF PHY-1125915. M.S. is grateful to KITP where this paper was completed for kind hospitality extended to him during the Workshop ``Resurgent Asymptotics in Physics and Mathematics.''   The work of S.B and K.K. is  supported by the INFN special 
project grant, ``GAST (Gauge and String Theory)".   M.S. thanks INFN, Pisa and University of Pisa for warm hospitality received during his visit to Pisa, when the collaboration on the present work started. 

%\newpage


\begin{thebibliography}{99}

\bibitem{th}
G. 't Hooft, {\em Naturalness, Chiral Symmetry, and Spontaneous Chiral Symmetry
Breaking}, in {\sl Recent Developments In Gauge Theories},  Eds. G. 't Hooft,
C. Itzykson, A. Jaffe, H. Lehmann, P. K. Mitter, I. M. Singer and R. Stora,
(Plenum Press, New York, 1980) [Reprinted in {\sl Dynamical Symmetry Breaking},
Ed. E. Farhi et al. (World Scientific, Singapore, 1982) p. 345 and in G. 't
Hooft, {\sl Under the Spell of the Gauge Principle}, (World Scientific, Singapore,
1994), p. 352].

\bibitem{Raby}
Stuart Raby, Savas Dimopoulos, and Leonard Susskind, 
{\em Tumbling Gauge Theories},
Nucl. Phys. {\bf B169}, 373  (1980).

\bibitem{Shrock}
T.~Appelquist, A.~G.~Cohen, M.~Schmaltz and R.~Shrock,
{\em New constraints on chiral gauge theories,}
  Phys.\ Lett.\ B {\bf 459}, 235 (1999)
 % doi:10.1016/S0370-2693(99)00616-4
  [hep-th/9904172];
  %%CITATION = doi:10.1016/S0370-2693(99)00616-4;%%
  T.~Appelquist, Z.~y.~Duan and F.~Sannino,
  {\em Phases of chiral gauge theories,}
  Phys.\ Rev.\ D {\bf 61}, 125009 (2000)
  %doi:10.1103/PhysRevD.61.125009
  [hep-ph/0001043];
  Y.~L.~Shi and R.~Shrock,
  {\em$A_k \bar F$ chiral gauge theories,}
  Phys.\ Rev.\ D {\bf 92},105032 (2015)
  %doi:10.1103/PhysRevD.92.105032
  [arXiv:1510.07663 [hep-th]];
  %%CITATION = doi:10.1103/PhysRevD.92.105032;%%
  Y.~L.~Shi and R.~Shrock,
  {\em Renormalization-Group Evolution and Nonperturbative Behavior of Chiral Gauge Theories with Fermions in Higher-Dimensional Representations,}
  Phys.\ Rev.\ D {\bf 92}, 125009 (2015)
  [arXiv:1509.08501 [hep-th]]. 

  
  \bibitem{Poppitz}
  E.~Poppitz and Y.~Shang,
{\em Chiral Lattice Gauge Theories Via Mirror-Fermion Decoupling: A Mission (im)Possible?},
  Int.\ J.\ Mod.\ Phys.\ A {\bf 25}, 2761 (2010)
 % doi:10.1142/S0217751X10049852
  [arXiv:1003.5896 [hep-lat]].
  %%CITATION = doi:10.1142/S0217751X10049852;%%
  %18 citations counted in INSPIRE as of 24 Jan 2018

\bibitem{AS}
A.~Armoni and M.~Shifman,
{\em A Chiral SU(N) Gauge Theory Planar Equivalent to Super-Yang-Mills,}
  Phys.\ Rev.\ D {\bf 85}, 105003 (2012)
  %doi:10.1103/PhysRevD.85.105003
  [arXiv:1202.1657 [hep-th]].
  %%CITATION = doi:10.1103/PhysRevD.85.105003;%%
  
  \bibitem{Fradkin}
  E.~H.~Fradkin and S.~H.~Shenker,
{\em Phase Diagrams of Lattice Gauge Theories with Higgs Fields,}
  Phys.\ Rev.\ D {\bf 19} (1979) 3682.
%  doi:10.1103/PhysRevD.19.3682
  %%CITATION = doi:10.1103/PhysRevD.19.3682;%%
  
  \bibitem{shify}
  M.~Shifman and A.~Yung,
  {\em Fradkin Shenker continuity and instead-of-confinement phase,}
  Mod.\ Phys.\ Lett.\ A {\bf 32}, no. 30, 1750159 (2017)
  %doi:10.1142/S0217732317501590
  [arXiv:1704.06201 [hep-ph]].
  %%CITATION = doi:10.1142/S0217732317501590;%%
  
  \bibitem{SW}
N.~Seiberg and E.~Witten,
{\em  Electric - magnetic duality, monopole condensation, and confinement
  in N=2 supersymmetric Yang-Mills theory}',
{Nucl.Phys.~B426,~19~(1994)},
hep-th/9407087;
%
%%CITATION = HEP-TH/9408099;%%
N.~Seiberg and E.~Witten,
{\em Monopoles, duality and chiral symmetry breaking in N=2
  supersymmetric QCD},
Nucl.~Phys.~B431,~484~(1994),
hep-th/9408099.

  \bibitem{unsal}
  M.~\"Unsal,
  {\em Magnetic bion condensation: A New mechanism of confinement and mass gap in four dimensions,}
  Phys.\ Rev.\ D {\bf 80}, 065001 (2009)
 % doi:10.1103/PhysRevD.80.065001
  [arXiv:0709.3269 [hep-th]];
  %%CITATION = doi:10.1103/PhysRevD.80.065001;%%
  %120 citations counted in INSPIRE as of 05 Dec 2017
M.~\"Unsal and L.~G.~Yaffe,
{\em Center-stabilized Yang-Mills theory: Confinement and large-$N$ volume independence,}
  Phys.\ Rev.\ D {\bf 78}, 065035 (2008)
 % doi:10.1103/PhysRevD.78.065035
  [arXiv:0803.0344 [hep-th]];
  %%CITATION = doi:10.1103/PhysRevD.78.065035;%%
  %189 citations counted in INSPIRE as of 05 Dec 2017
  M.~Shifman and M.~\"Unsal,
 {\em QCD-like Theories on $R(3) \times S_1$: A Smooth Journey from Small to Large $r(S_1)$ with Double-Trace Deformations,}
  Phys.\ Rev.\ D {\bf 78}, 065004 (2008)
 % doi:10.1103/PhysRevD.78.065004
  [arXiv:0802.1232 [hep-th]].
  %%CITATION = doi:10.1103/PhysRevD.78.065004;%%
  %70 citations counted in INSPIRE as of 05 Dec 2017
  
  \bibitem{aaunsal}
   A.~Armoni, M.~Shifman and M.~\"Unsal,
{\em Planar Limit of Orientifold Field Theories and Emergent Center Symmetry,}
  Phys.\ Rev.\ D {\bf 77}, 045012 (2008)
 % doi:10.1103/PhysRevD.77.045012
  [arXiv:0712.0672 [hep-th]].
  %%CITATION = doi:10.1103/PhysRevD.77.045012;%%
  %16 citations counted in INSPIRE as of 05 Dec 2017
  
 \bibitem{unsalms}
  M.~Shifman and M.~\"Unsal,
{\em On Yang-Mills Theories with Chiral Matter at Strong Coupling,}
  Phys.\ Rev.\ D {\bf 79}, 105010 (2009)
 % doi:10.1103/PhysRevD.79.105010
  [arXiv:0808.2485 [hep-th]].
  %%CITATION = doi:10.1103/PhysRevD.79.105010;%%
  %19 citations counted in INSPIRE as of 05 Dec 2017
  
  \bibitem{CSU}
  A. Cherman, M. Shifman and M. \"Unsal, to be published.

\end{thebibliography}
\end{document}